\documentclass[12pt]{article}
\usepackage{amsfonts}
\usepackage{amsmath}
\usepackage{amssymb}
\usepackage{graphicx}
\usepackage{textcomp}
\begin{document}
\sf
\begin{center}
   \vskip 2em
    {\LARGE \sf 
Brownian motion meets Riemann curvature 
    }\\
 \vskip 2em

 {\large \sf  Pavel Castro Villarreal\footnote{E-mail: pcastrov@unach.mx} \\[0.5em]}
\em{ Centro de Estudios en F\'{\i}sica y Matem\'aticas B\'asicas y Aplicadas, \\Universidad Aut\'onoma de Chiapas, 
 C.P. 29000, 1428 Tuxtla Guti\'errez, Chiapas, M\'EXICO\\[1em]}
\end{center}
 \vskip 1em
\begin{abstract}

The general covariance of the diffusion equation is exploited  in order to explore the curvature effects appearing on brownian motion over a d-dimensional curved manifold. We use the local frame defined by the so called Riemann normal coordinates to derive a general formula for the mean-square geodesic distance (MSD) at the short-time regime. This formula is written in terms of $O(d)$ invariants that depend on the Riemann curvature tensor. We study the n-dimensional sphere case to validate these results. We also show that the  diffusion for positive constant curvature is slower than  the diffusion in a plane space, while the diffusion for negative constant curvature turns out to be faster.  Finally the two-dimensional case is emphasized, as it is  relevant for the single particle diffusion on biomembranes.

\end{abstract}

\section{Introduction}

The brownian motion phenomenon occurs in a wide diversity of physical areas; from colloidal physics to quantum gravity and biophysics (see for instance \cite{Colloid},\cite{Duplantier1} and \cite{Almeida}). In the last decade, motivated by problems  in biophysics  (see \cite{Weiss, Ivo}), an intense activity has emerged on the  study of diffusion processes on curved manifolds.  For instance,  the transport phenomena occurring in cell membranes is an interesting and complex problem. In particular, the random motion of  a single integral protein  or lipid in a cell membrane is difficult to realise, mainly because of the  interactions and obstacles with the remainder components of the cell.  In addition to this difficulty, the thermal fluctuations produce shape undulations on the curved membrane \cite{Seifert, Naji, Seifert2}.  By simplifying this problem to the study of free diffusion on a cell membrane, considered as a regular and continuos two-dimensional surface,  Smolouchovski's  diffusion equation has been proposed in \cite{Aizenbud, Gustaffson}. Explicit formulas for constant mean  and gaussian curvatures have been presented at \cite{H} and \cite{K}, respectively.  Nevertheless, the analytical issues that appear on the general surface case, have motivated the incorporation of novel computer simulations \cite{Holyst, Christensen} (see also \cite{Ivo} for a related work).

From a theoretical stand point,  the brownian motion can be used to probe the geometry of  the manifold in the spirit  of Kac's famous question: {\it Can one hear the shape of a drum?} \cite{Kac},  and vice versa, the geometry will cause a change in the standard randomness of the particle motion \cite{Einstein}. It is then imminent to analyse a quantitative contribution  coming from the geometry, and to be specific, how the curvature  of the manifold affects the motion. These questions have been arisen in \cite{K}, where the local concentration of a diffused substance is obtained in terms of the local curvature. Furthermore,  a detailed study of the way on how the mean-values are affected by the curvature for two-dimensional manifolds is presented in \cite{Faraudo},  with a special emphasis on developable and isotropic surfaces.  Also, the study of particular cases presented in \cite{Tomoyoshi} suggests that for negative  Gaussian curvature, the diffusion accelerates; whereas, the diffusion decelerates for surfaces with positive Gaussian curvature.  
We should point out that, although with different formalism, these same questions have been posted in  \cite{Fabrice}.

In this work we explore the curvature effects on the brownian motion when the particle movement takes  place on a  d-dimensional  riemannian manifold. These effects manifest differently for different physical observables. Here,  we will use the geodesic distance as the displacement of the particle, but some other observables can also be defined \cite{Gustaffson, Holyst, Faraudo}, where either intrinsic or extrinsic properties of the manifold can be probed \cite{Castro}.  Furthermore, we will take advantage of the general covariance of the diffusion equation to use a special frame defined by the so called Riemann Normal Coordinates (RNC) \cite{Eisenhart}. In this frame we will  compute curvature corrections for the mean-square geodesic distance. In particular, we use a technique developed in \cite{Denjoe}, originally used  to compute curvature corrections that appear in effective actions of field theory on curved spaces (see for instance \cite{Denjoe1, Denjoe2}). Related work concerning the RNC can be found at \cite{Muller, Hatzinikitas}.  It is remarkable  that, using these coordinates, the geodesic curves on the manifold look like stright lines and therefore, the square geodesic distance $s^{2}$ will have the same structure as the square distance in an euclidean space: $s^{2}=\delta_{ab}y^{a}y^{b}$ (where $y^{a}$ are the RNC). As it is shown in this paper, the mean-square geodesic distance is clearly isometric, this is because it  only depends  on $O(d)$ invariant combinations of the Riemann curvature tensor.  

This paper is organized as follows: In section 2  we summarize the geometrical concepts used  to approach the brownian motion. In particular, we introduce the  frame defined by the Riemann Normal Coordinates.  In section 3, we present the diffusion equation on curved manifolds and we derive general remarks concerning the short time regime. In section 4, we focus on the computation of the mean-square geodesic displacement using  RNC. In particular, we study the curvature effects for the manifolds with constant curvature and the two-dimensional case. The $n$-dimensional sphere is also explore to validate our results. Finally, in section 5 we sintetize the main conclusions and perspectives of this work.

\section{Geometrical preliminaries and notation}

In this section we review the preliminary notions about manifolds and riemannian geometries  (following \cite{Nakahara}) needed to describe the brownian motion.  Let us call $\mathbb{M}$ a $d$-dimensional manifold with local coordinates $\varphi:U\subset \mathbb{M}\to \mathbb{R}^{d}$, where $U$ is a local neighbourhood and $\mathbb{R}^{d}$ is the $d$-dimensional euclidean space. As a consequence of the differentiablity of the map $\varphi$, the set $U$ is locally diffeomorphic to a piece of the euclidean space. The local coordinates are also denoted by $\varphi(p)=(x^{1},\cdots,x^{d})$ where $p\in\mathbb{M}$.  For each point $p$ on the manifold we associate a vector space called the tangent space ${\rm T}_{p}\mathbb{M}$, whose elements are denoted by capital letters  $X, Y, Z, \cdots$. 

We are interested in manifolds endowed with a {\it riemannian metric}. If $g_{p}:{\rm T}_{p}\mathbb{M}\times {\rm T}_{p}\mathbb{M}\to\mathbb{R}$ denotes a riemannian metric, we write 
\begin{eqnarray}
g_{p}=g_{ab}~dx^{a}\otimes dx^{b},
\end{eqnarray}
where $\left\{dx^{a}\right\}$ constitute a one-form basis of the dual tangent space ${\rm T}^{*}_{p}\mathbb{M}$ and $g_{ab}$ is the metric tensor. The Einstein summation rule is adopted for the repeated indexes.  The knowledge of the metric tensor components allow us to compute further geometrical quantities. The geometrical meaning of how the manifold is curved is defined in terms of  the torsion tensor and the Riemann curvature tensor \cite{Nakahara},
\begin{eqnarray}
T\left(X,Y\right)&\equiv& \nabla_{X}Y-\nabla_{Y}X-\left[X,Y\right],\nonumber\\
R\left(X,Y,Z\right)&\equiv& \nabla_{X}\nabla_{Y}Z-\nabla_{Y}\nabla_{X}Z-\nabla_{\left[X,Y\right]}Z
\label{Def}
\label{curvature}
\end{eqnarray}
where $\nabla$ is the affine connection. For our purposes, we  will use the Levi-Civita connection, which is the only one compatible with the metric. Using the coordinate basis $\left\{e_{a}\right\}\equiv\left\{\partial_{a}\right\}$  of the tangent space ${\rm T}_{p}\mathbb{M}$,  the affine connection defines the components $\Gamma^{a}_{~bc}$ by
\begin{eqnarray}
\nabla_{a}e_{b}\equiv\nabla_{e_{a}}e_{b}=\Gamma^{c}_{~ab}e_{c},
\end{eqnarray} 
where $\nabla_{a}$ stands for the covariant derivative and for the Christoffel symbols $\Gamma^{a}_{~bc}$ \footnote{$\Gamma^{a}_{~bc}=\frac{1}{2}g^{ae}\left(\partial_{b}g_{ec}+\partial_{c}g_{eb}-\partial_{e}g_{bc}\right)$}.  The torsion  is a (1,2) type tensor $T=T^{a}_{~bc}~e_{a}\otimes dx^{b}\otimes dx^{c}$
whereas the Riemann curvature is a (1,3) type one $R=R^{a}_{~bcd}~e_{a}\otimes dx^{b}\otimes dx^{c}\otimes dx^ {d}$. The components of these tensors  are given in terms of  the Chrystoffel symbols \footnote{$T^{a}_{ ~bc}=\Gamma^{a}_{~bc}-\Gamma^{a}_{~cb}$, and $R^{a}_{~bcd}=\partial_{c}\Gamma^{a}_{~db}-\partial_{d}\Gamma^{a}_{~cb}+\Gamma^{e}_{~db}\Gamma^{a}_{~ce}-\Gamma^{e}_{~cb}\Gamma^{a}_{~de}$.} . Clearly, the manifold is free torsion for the Levy-Civita connection, because the Christoffel symbols are symmetric.  The quantity $R_{abcd}\equiv g_{af}R^{f}_{~bcd}$ satisfies the following useful identities $R_{abcd}=-R_{bacd}=-R_{abdc}=R_{cdab}$. Using the Riemann curvature components we can defined the Ricci tensor $R_{ab}\equiv R^{c}_{~acb}$ and the scalar curvature $R_{g}=g^{ab}R_{ab}$.

There is an important device introduced by Riemann, nowadays called the Riemann Normal Coordinates \cite{Eisenhart}. This coordinate system can be defined by mapping a point $p$ on the manifold to the origin of $\mathbb{R}^{d}$  and the following conditions
\begin{eqnarray}
g_{ab}(0)=\delta_{ab}, ~~~~~~~y^{a}g_{ab}(y)=y^{a}\delta_{ab}.
\label{defRNC}
\end{eqnarray}
As it is point out at \cite{Muller},  the second condition is equivalent to the following gauge condition  on the affine connection (see appendix A)
\begin{eqnarray}
y^{a}y^{b}\Gamma^{c}_{~ab}(y)=0.
\label{normality}
\end{eqnarray}
Furthermore, in RNC the Taylor coefficients of the metric tensor can be found in terms of the covariant derivatives of the Riemann curvature tensor \cite{Denjoe, Muller, Hatzinikitas} as
 \begin{eqnarray}
g_{ab}\left(y\right)&=&\delta_{ab}+\frac{1}{3}R_{acdb}\left(0\right)y^{c}y^{d}+\frac{1}{6}\nabla_{e}R_{acdb}\left(0\right) y^{e}y^{c}y^{d}+\frac{2}{45}R_{acdf}\left(0\right)R^{f}_{~ghb}\left(0\right)y^{c}y^{d}y^{g}y^{h}\nonumber\\&+&\frac{1}{20}\nabla_{e}\nabla_{f}R_{acdb}\left(0\right)y^{e}y^{f}y^{c}y^{d}+\cdot\cdot\cdot\nonumber,
\label{MetricRNC}
\end{eqnarray}
where $y$ denotes $\varphi(q)$, the RNC, and $q$ belong to the same patch of $p$. See appendix A for a derivation of this series expansion.  Using these coordinates, the geodesic curves look like straight lines passing through the point $p$. Indeed, using the gauge condition (\ref{normality}) and the geodesic equation it is easy to figure that out. Therefore, the geodesic curve can be written as $y^{a}\left(s\right)=\xi^{a}s$, where $s$ is the geodesic distance and $\xi^{a}$ are constants \cite{Hatzinikitas}. Furthermore, as the geodesic curve is parametrized by the arc-lenght, the coeffitients $\xi^{a}$ satisfy $g_{ab}\xi^{a}\xi^{b}=1$, so the square geodesic distance is given by
\begin{eqnarray}
s^{2}=g_{ab}y^{a}y^{b}=\delta_{ab}y^{a}y^{b},
\label{sdispl}
\end{eqnarray}
where the last equality comes  from the conditions of the RNC (\ref{defRNC}). This equation is remarkable, because the geodesic distance has the same form as in the euclidean geometry.

\section{Diffusion and  geometry}

Here, we introduce the simplest model for the brownian motion of a free particle, which takes place on a d-dimensional riemannian geometry.  This is a direct generalization of the brownian motion on euclidean spaces, which basically consists on replacing the euclidean laplacian by the Laplace-Beltrami operator  in the diffusion equation \cite{Aizenbud, Gustaffson}
\begin{eqnarray}
\frac{\partial P\left(x,x^{\prime},t\right)}{\partial t}&=&D\Delta_{g}P\left(x,x^{\prime},t\right),\nonumber\\
P\left(x,x^{\prime},0\right)&=&\frac{1}{\sqrt{g}}\delta^{\left(d\right)}\left(x-x^{\prime}\right).
\label{difeq}
\end{eqnarray}
Here, $P\left(x,x^{\prime},t\right)dv$ is the probability to find the diffusing particle in the volume element $dv=\sqrt{g}d^{d}x$ when the particle  started to move at $x^{\prime}$.  The probability distribution $P\left(x,x^{\prime},t\right)$ is normalized with respect to the volume $v$ of the manifold and $D$ is the diffusion coefficient.  The operator $\Delta_{g}$ is the Laplace-Beltrami operator, which is defined by
\begin{eqnarray}
\Delta_{g}\cdot=\frac{1}{\sqrt{g}}\partial_{a}\left(\sqrt{g}g^{ab}\partial_{b}~\cdot~\right),
\label{L-Bop}
\end{eqnarray}
where $g=\det\left( g_{ab}\right)$.  We should point out that when the manifold is not compact, we will require for the probability and all its partial derivatives to  vanish at the boundary.  The formal solution of the diffusion equation (\ref{difeq})  on curved manifolds (see  \cite{DeWitt, Denjoe2}) is given in terms of the Minakshisundaram-Pleijel coeffients, which depend on both $x$ and $x^{\prime}$. This solution  has already  been used in order to describe the concentration of a diffused substance over curved manifolds  in the limit when $x\to x^{\prime}$ \cite{K}.

Once we have a probability distribution, we want to look at the mean-values of physical observables (for example, the mean-square displacement) in order to get information of the brownian motion.  For scalar functions $\Omega$  (defined on the manifold), the expectation values are defined in the standard fashion
\begin{eqnarray}
\left<\Omega\left(x\right)\right>_{t}=\int_{\mathbb{M}} dv~\Omega\left(x\right)P\left(x,x^{\prime},t\right).
\end{eqnarray}
Note that $\left<\Omega\left(x\right)\right>_{t}$ also depends on the initial point $x^{\prime}$.
In principle, it is posible to evaluate the expectation values using the formal solution mentioned above. However, using this procedure it may be very involve because the Minakshisundaram-Pleijel coeffients, as far as we know, are not known for points $x$ different from $x^{\prime}$. Here, we use a different strategy, which will be applied only for physical observables well behaved under actions of $\Delta_{g}$.

\subsection{General remark on the short time asymptotics}
Clearly, when $\Omega(x)$ is well behaved under any number of actions of the Laplace-Beltrami operator, its expectation value can be expanded in Taylor series in the variable $t$. The $k-{\rm th}$ derivative  of $\left<\Omega\left(x\right)\right>_{t}$ at $t=0$ can be computed as follows. First, let us compute the first derivative using the diffusion equation, 
\begin{eqnarray}
\frac{\partial\left<\Omega\left(x\right)\right>_{t}}{\partial t}&=&D\int_{\mathbb{M}} dv~\Omega\left(x\right)\Delta_{g}P\left(x,x^{\prime},t\right)\nonumber\\
&=&D\int_{\mathbb{M}} dv~\Delta_{g}\Omega\left(x\right)P\left(x,x^{\prime},t\right)+\int_{\mathbb{M}} dv ~\nabla_{a}J^{a},
\end{eqnarray}
where $J^{a}$ is a boundary current given by $J^{a}=g^{ab}\partial_{b}\left(\Omega P\right)$. Using this procedure it is posible to compute the $k-{\rm th}$ Taylor coeffitient by 
\begin{eqnarray}
\left.\frac{\partial^{k} \left<\Omega\left(x\right)\right>_{t}}{\partial t^{k}}\right |_{t=0}=D^{k}\Delta^{k}_{g}\Omega\left(x^{\prime}\right),
\end{eqnarray}
here we dropped all the boundary terms because  the probability and its derivatives vanish there. The expectation value for our physical observable is then given by the formal series \cite{Grygorian}
\begin{eqnarray}
\left<\Omega\left(x\right)\right>_{t}=\left .\left[1-\frac{1}{1!}t\mathcal{H}+\frac{1}{2!}t^{2}\mathcal{H}^{2}+\cdot\cdot\cdot\right]\Omega\left(x\right)\right |_{x=x^{\prime}}
\label{computation}
\end{eqnarray}
where $\mathcal{H}=-D\Delta_{g}$.   This expression is very useful to access the short time regime of the brownian motion. Indeed, given a physical observable which is well behaved under the actions of the Laplace-Beltrami operator, its mean-value at the short-time regime can be calculated using (\ref{computation}).   In particular, for the plane case $\mathbb{R}^{d}$ we want  to know the mean-square displacement $\left|{\bf x}\right|^{2}$ (where ${\bf x}\in \mathbb{R}^{d}$ and we have chosen ${\bf x}^{\prime}=0$). In this case, the Laplace-Beltrami reduces simply to the laplacian $\partial^{a}\partial_{a}$.  Appliying formula (\ref{computation}) we get the standard kinematical Einstein relation for the mean-square displacement:  $\left<\left|{\bf x}\right|^{2}\right>_{t}=2dDt$. Observe,we can not compute the mean value for the displacement $\left|{\bf x}\right|$ by this method because it is not  well behaved under the actions of the laplacian at ${\bf x}=0$.

\section{The mean values of $s^{2}$ and short time asymptotics}

On curved manifold there are several quantities that can be used to described the brownian motion. For instance, in \cite{Holyst} the particle position is given in terms of the parametrization of a manifold embedded in the ambient space $\mathbb{R}^{3}$.  For this case,   the displacement is given by the euclidean norm of the parametrization.  However, in \cite{Faraudo} the brownian displacement  is defined by the geodesic distance. In addition, using the Monge parametrization for a surface we can also define a projected displacement \cite{Gustaffson}.  Here, nevertheless, we will not  discern between these quantities, instead, we will stress the fact that  all of them represent different manifestations of the same phenomenon.  The analysis between these observables is beyond the scope  of this work. 

In this paper, we use the geodesic distance as the definition of the  displacement of the particle. As in the plane case, this quantity is rotational and traslational invariant. Furthermore, the geodesic distance is invariant under general coordinate transformations.  Therefore the mean-value of $s^{2}$ will be expected to be invariant under a general coordinate transformation. In what follows  we compute the expectation value of the square displacement using the Riemann normal frame centered at the point $p=\varphi^{-1}(0)$. Therefore the particle starts to move at the initial condition and the expectation value of $s^{2}$  can be computed using (\ref{computation}) 
\begin{eqnarray}
\left<s^{2}\right>_{t}=\sum^{\infty}_{k=1}\frac{G_{k}}{k!}\left(Dt\right)^{k}
\label{mean-squaregeo}
\end{eqnarray}
where the terms $G_{k}=\left. \Delta^{k}_{g}s^{2}\right |_{y=0}$ ($k=1,2,3,\dots$) are purely geometric factors. This factors can be computed explicitly using the technology of the RNC.  We will compute the first three factors, $G_{0}$, $G_{1}$ and $G_{3}$.  

Using the definition of the Laplace-Beltrami operator (\ref{L-Bop}) and $s^{2}=\delta_{ab}y^{a}y^{b}$,  it is not difficult to show that, for every coordinate $y$ on the manifold,
\begin{eqnarray}
\Delta_{g}s^{2}=2d+y\cdot\partial\left(\log g\right),
\label{LBonS}
\end{eqnarray}
 where $g$ is the determinant of the metric and $y\cdot\partial=y^{a}\partial_{a}$.  Clearly, when we evaluate at $y=0$ we get $G_{1}=2d$.  The factors $G_{2}$ and $G_{3}$ can be found using the result of the appendix B, summarized as follows, if $f$ is a diferentiable function on the manifold then the Laplace-Beltrami operator acting on $f$ at $y=0$  is simply  by $\left.\partial^{a}\partial_{a}f\right |_{y=0}$. Therefore $ G_{2}=\left.\partial^{2}\left(\Delta_{g}s^{2}\right)\right|_{y=0}$ and $G_{3}=\left.\partial^{2}\left(\Delta^{2}_{g}s^{2}\right)\right|_{y=0}$. So, we need at least the second order  Taylor expansion of $\Delta_{g}s^{2}$ and $\Delta^{2}_{g}s^{2}$. Using equation (\ref{LBonS}) and the Taylor expansion of $\log g$ (see appendix A), we obtain the second order of 
 $\Delta_{g}s^{2}$, 
 \begin{eqnarray}
\Delta_{g}s^{2}=2d-\frac{2}{3}R_{ab}\left(0\right)y^{a}y^{b}+O(y^{3}).
\end{eqnarray}
Therefore, the second geometric factor is $G_{2}=-\frac{4}{3}R_{g}$, where the Ricci scalar curvature is evaluated at $y=0$.  For the third gometric factor $G_{3}$, let  $\Delta_{g}$ act on  equation (\ref{LBonS}), 
\begin{eqnarray}
\Delta^{2}_{g}s^{2}=\frac{1}{2}\left(\partial_{a}\log g\right)g^{ab}\partial_{b}\left(y\cdot\partial\log g\right)+\left(\partial_{a}g^{ab}\right)\partial_{b}\left(y\cdot\partial\log g\right)+g^{ab}\partial_{a}\partial_{b}\left(y\cdot\partial\log g\right).
\end{eqnarray}
Using the perturbative expansion of the inverse metric $g^{ab}$ and the one of the $\log g$ it is not difficult to obtain the second order of $\Delta^{2}_{g}s^{2}$ (see appendix A). Therefore, by a straightforward calculation, we get
\begin{eqnarray}
G_{3}=\frac{8}{15}R^{ab}R_{ab}-\frac{16}{45}R^{abcd}\left(R_{dbca}+R_{dcba}\right)-\frac{16}{5}\left(\nabla^{a}\nabla^{b}+\frac{1}{2}g^{ab}\Delta_{g}\right)R_{ab}.
\end{eqnarray}
In general, for the $G_{k}$ factor we need to compute the second order pertubative expression for $\Delta^{k-1}_{g}s^{2}$.  

For the expectation value of $s^{2}$ at the short-time regime we have considered only the values $k=1,2,3$. Hence, this value can be written as
\begin{eqnarray}
\left<s^{2}\right>_{t}=2dDt-\frac{2}{3}R_{g}\left(Dt\right)^{2}&+&\frac{1}{3!}\left[\frac{8}{15}R^{ab}R_{ab}\right.-\left.\frac{16}{45}R^{abcd}\left(R_{dbca}+R_{dcba}\right)\right.\nonumber\\
&&~~~~~~~~~~~~~~~~~-\left.\frac{16}{5}\left(\nabla^{a}\nabla^{b}+\frac{1}{2}g^{ab}\Delta_{g}\right)R_{ab}\right]\left(Dt\right)^{3}+\cdot\cdot\cdot
\label{mean-square}
\end{eqnarray}
As we anticipate in the introduction, the mean-square geodesic distance (i.e. $\left<s^{2}\right>_{t}$) is deviated from the planar expression by terms which are $O(d)$ invariant as well as invariant under a general coordinate transformation. Clearly, for very short-times,  the particle movement is not affected by the curvature of the manifold and the standard mean-square displacement is recovered \cite{Einstein}.  This follows from the very nature of the manifold (every local neighbourhood looks like a piece of an euclidean space). However, as the particle explores away from the local boundaries, the movement is affected by the curvature of the manifold.  Additionally, the curvature corrections to the planar expression are isometric. Therefore, every manifold which is isometric to an euclidean space, will have null curvature effects for this observable. In particular, this is the case for the developable surfaces \cite{Faraudo}.

\subsection{Example: the brownian motion on the $n$-sphere}

Here, the idea is to perform a cross-check calculation in order to validate equation (\ref{mean-square}). In particular,  we compute the mean-square geodesic distance at the short-time regime,   when the brownian motion takes place on a n-dimensional sphere. The hypersphere $S^{n}$ of radius $R$ is defined by
\begin{eqnarray}
S^{n}=\left\{{\bf x}\in \mathbb{R}^{n+1}:{\bf x}^{2}=R^{2}\right\}.
\end{eqnarray} 
This manifold can be parametrized using the local coordinates $\left(\theta_{0},\theta_{1},\cdots,\theta_{n-1}\right)$, where $\theta_{0}$ takes values in $[0,\pi]$ whereas the remainder coordinates in $[0,2\pi)$. In this example, we use a parametrization concerning to the ambient space ${\mathbb{R}^{n+1}}$ by using the functions ${\bf x}=(x_{0},\cdots,x_{n})$ given by
\begin{eqnarray}
x_{0}&=&R\cos\theta_{0}\nonumber\\
x_{1}&=&R\sin\theta_{0}\cos\theta_{1}\nonumber\\
\vdots&&\nonumber\\
x_{n}&=&R\sin\theta_{0}\sin\theta_{1}\cdots\sin\theta_{n-1}.
\end{eqnarray}
Clearly, ${\bf x}=R{\bf n}$, where ${\bf n}$ is the unit normal vector pointing
outward from hypersurface of  $S^{n}$. The metric tensor can be computed using $g_{ab}=\partial_{a}{\bf x}\cdot\partial_{b}{\bf x}$ and it can be written in a matrix form as
\begin{eqnarray}
g_{ab}=\left(\begin{array}{ccccc}
1& & & & {\Large 0}\\
&\sin^{2}\theta_{0}& & &\\
& &\sin^{2}\theta_{0}\sin^{2}\theta_{1} &&\\
& & & \ddots&\\
{\Huge 0}& && &\sin^{2}\theta_{0}\cdots\sin^{2}\theta_{n-1}
\end{array}
\right).
\end{eqnarray}
The square root of the metric tensor determinant is $\sqrt{g}=R^{n}\sin^{n-1}\theta_{0}\sin^{n-2}\theta_{1}\cdots\sin\theta_{n-1}$. It is re\-mar\-ka\-ble that using extrinsic geometry we  can easily compute the Riemann curvature tensor. Indeed,  using the  Gauss-Codazzi equations,  $R_{abcd}=K_{ac}K_{bd}-K_{ad}K_{bc}$, where $K_{ab}=\partial_{a}{\bf x}\cdot\partial_{b}{\bf n}$ are the components of the second fundamental form \cite{spivak}, the Riemann curvature tensor is 
\begin{eqnarray}
R_{abcd}=\frac{1}{R^{2}}\left(g_{ac}g_{bd}-g_{ad}g_{bc}\right).
\label{RiemannSphere}
\end{eqnarray}
The Ricci tensor  and the scalar curvature are then given by
\begin{eqnarray}
R_{ab}=\frac{n-1}{R^{2}}g_{ab},~~~~~~~~~
R_{g}=\frac{n\left(n-1\right)}{R^{2}},
\label{RicciScalar}
\end{eqnarray}
respectively. Substituting equations (\ref{RiemannSphere}) and (\ref{RicciScalar}) into the expressions of $G_{1}$, $G_{2}$, and $G_{3}$, obtained above,   we find
\begin{eqnarray}
G_{1}=2n,~~~~~~G_{2}=-\frac{4}{3}\frac{n(n-1)}{R^{2}},~~~~~~
G_{3}=\frac{8}{15}\frac{n(n-1)(n-3)}{R^{4}}
\label{Gs}
\end{eqnarray}
Now, in order to perform an indepedient calculation we compute the mean-square geodesic distance using the equation (\ref{computation}). For practical purpose, we use the geodesic curve starting at the  north pole defined as the point ${\bf x}=(1,0,\cdots,0)$ of $S^{n}$. It is not difficult to show that $\theta_{0}=\frac{1}{R}s$ and $\theta_{j}=0$ (with $j=1,\cdots,n-1$) is this geodesic curve. The geodesic distance is simply given by $s=R\theta_{0}$.

In this case, the Laplace-Beltrami operator on $S^{n}$ can be splitted as
\begin{eqnarray}
\Delta_{g}=\frac{1}{R^{2}\sin^{n-1}\theta_{0}}\partial_{0}\sin^{n-1}\theta_{0}\partial_{0}+\sum^{n-1}_{i,j=1}\frac{1}{\sqrt{g}}\partial_{i}\sqrt{g}g^{ij}\partial_{j},
\label{LBonSph}
\end{eqnarray}
where $\partial_{0}\equiv\frac{\partial}{\partial\theta_{0}}$ and $\partial_{j}\equiv\frac{\partial}{\partial\theta_{j}}$. Clearly, the actions of the Laplace-Beltrami operator on $s^{2}$  will involve only the first term of (\ref{LBonSph}).  The first, second and third action of $\Delta_{g}$ on $s^{2}$ are given by
\begin{eqnarray}
\Delta_{g}s^{2}&=&2(n-1)\theta_{0}\cot\theta_{0}+2,\nonumber\\
\Delta^{2}_{g}s^{2}&=&-\frac{2(n-1)}{R^{2}}\left[n-1+(n-3)(1+\cot^{2}\theta_{0})\left(\theta_{0}\cot\theta_{0}-1\right)\right],\nonumber\\
\Delta^{3}_{g}s^{2}&=&-\frac{2(n-1)(n-3)}{R^{4}}\left\{(n-1)\cot\theta_{0}(1+\cot^{2}\theta_{0})\left[\cot\theta_{0}(3-\theta_{0}\cot\theta_{0})-\theta_{0}(1+\cot^{2}\theta_{0})\right]\right.\nonumber\\
&+&\left.4(1+\cot^{2}\theta_{0})\left[(\theta_{0}\cot\theta_{0}-1)(2\cot^{2}\theta_{0}+1)-\cot\theta_{0}(\cot\theta_{0}-\theta_{0}(1+\cot^{2}\theta_{0}))\right]
\right\}.\nonumber\\
\end{eqnarray}
Therefore taking the limit when $\theta_{0}\to 0$ we get (\ref{Gs}).  The mean-square displacement is then given by
\begin{eqnarray}
\left<s^{2}\right>_{t}=2nDt-\frac{2}{3}\frac{n(n-1)}{R^{2}} \left(Dt\right)^{2}+\frac{4}{45}\frac{n(n-1)(n-3)}{R^{4}}\left(Dt\right)^{3}+\cdots
\end{eqnarray}
This equation is the desired result (\ref{mean-square}) for the particular case of the $n$-sphere. 

\subsection{The constant curvature spaces}

\begin{figure}[h]
\begin{center}
\includegraphics[width=8cm]{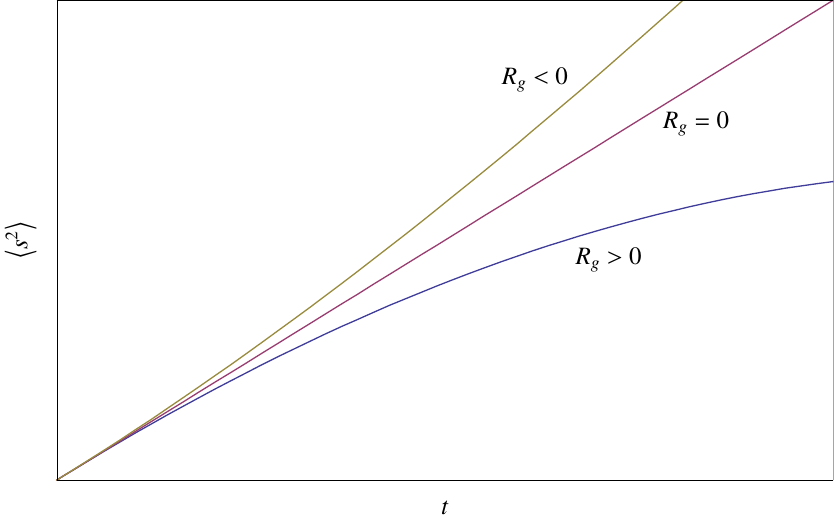}
\caption{{\small  The time evolution of the MSD  for the spherical $R_{g}>0$, hyperbolic $R_{g}<0$ and euclidean spaces $R_{g}=0$. }}
\label{fig1}
\end{center}
\end{figure}

For the constant curvature manifolds, the Riemann curvature tensor is given by \cite{spivak}
\begin{eqnarray}
R_{abcd}=\frac{R_{g}}{d(d-1)}(g_{ac}g_{bd}-g_{ad}g_{bc}),
\end{eqnarray}
where $R_{g}$ is a constant that can be either positive, negative or zero. It is known \cite{spivak} that the only three solutions for constant curvature are the d-dimensional sphere when $R_{g}>0$,  the d-dimensional hyperboloid when  $R_{g}<0$ and the euclidean space for $R_{g}=0$. For these cases, the mean-square geodesic distance is given by 
\begin{eqnarray}
\left<s^{2}\right>_{t}=2dDt-\frac{2}{3}R_{g}\left(Dt\right)^{2}+\frac{4}{45}\frac{d-3}{d(d-1)}R^{2}_{g}\left(Dt\right)^{3}+\cdots .
\label{MSC}
\end{eqnarray}
For theses cases, the curvature effects depends on the sign of the curvature $R_{g}$. For the hyperbolic spaces, the MSD deviates from the planar result by an increasing monotonic function (in time) whereas for the spherical space by a decreasing monotonic function. In other words, the geometry of the space affects the brownian motion in such a way that the  particle`s diffusion is accelerated for $R_{g}<0$ or decelerated for $R_{g}>0$; see figure (\ref{fig1}). These same results were also observed in the two-dimensional cases  in  \cite{Tomoyoshi}.

\subsection{The two-dimensional  case}

\begin{figure}[here]
\begin{center}
\includegraphics[width=8cm]{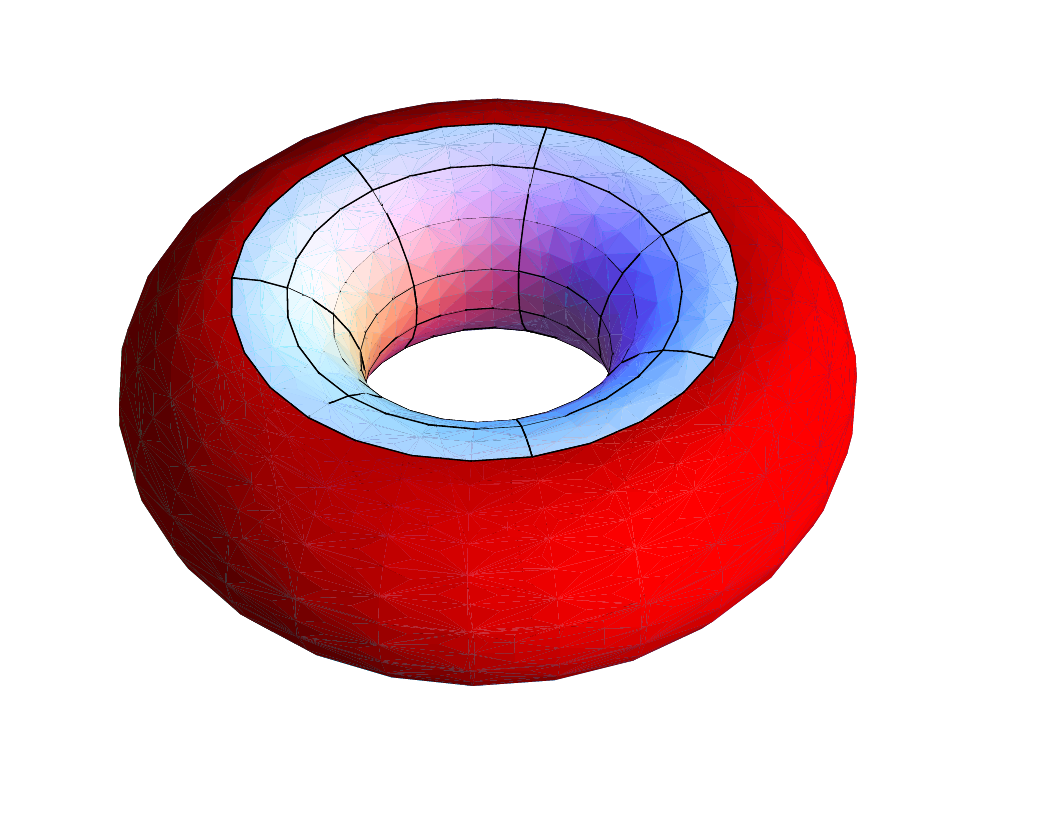}
\caption{{\small   A Torus divided by three regions, where $K>0$ , $K<0$ (i.e the area gridded) and $K=0$ (parallels circular curves where $\theta=\pi/2$ and $\theta=3\pi/2$). }}
\label{Figure}
\end{center}
\end{figure}
\begin{figure}[here]
\begin{center}
\includegraphics[width=8cm]{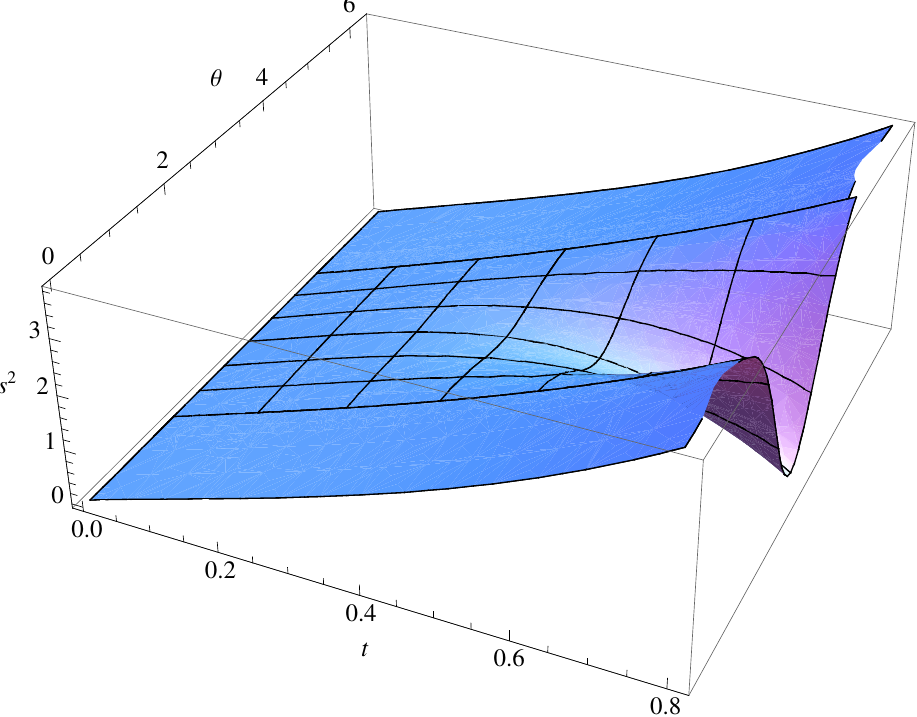}
\caption{{\small   The time evolution of $\left<s^{2}\right>_{t}$ versus $t$ and $\theta$, where the area gridded  represents the region where $K<0$. }}
\label{Figure2}
\end{center}
\end{figure}

The two-dimensional case is the most relevant one for the diffusion on biological membranes. For this case,  the Riemann curvature tensor is given by 
\begin{eqnarray}
R_{abcd}=K\left(g_{ac}g_{bd}-g_{ad}g_{bc}\right),
\end{eqnarray}
where $K=R_{g}/2$ is the Gaussian curvature  of the surface \cite{Nakahara}. Note that $K$ is not necessarily a constant.  The mean-square geodesic distance, at the short-time regime, for an arbitrary surface of Gaussian curvature $K$ is given by
\begin{eqnarray}
\left<s^{2}\right>_{t}=4Dt-\frac{4}{3}K\left(Dt\right)^{2}-\frac{8}{15}\left[\frac{1}{3}K^{2}
+2\Delta_{g}K\right]\left(Dt\right)^{3}+\cdot\cdot\cdot .
\label{MSD2}
\end{eqnarray}
In what follows, we derive few consequences of last equation when the diffusion takes place on a Torus. For this case, the metric is given by $ds^{2}=r^{2}d\theta^{2}+(a+r\cos\theta)^{2}d\varphi^{2}$ (with $0<\theta, \varphi<2\pi$), where $a$ and $r$ are the two radii.  The Gaussian curvature is given by 
\begin{eqnarray}
K=\frac{\cos\theta}{r(a+r\cos\theta)}.
\end{eqnarray}
At figure (\ref{Figure}) three regions on the Torus are shown. These regions are defined by the conditions $K>0$, $K<0$ and $K=0$, respectively\footnote{ Clearly,  the Gaussian curvature vanishes along the parallels ($\theta=\pi/2$ and $\theta=3\pi/2$). In the region given by $\pi/2<\theta<3\pi/2$, $K$ is negative whereas in the one  given by $0<\theta<\pi/2$ or $\pi/2<\theta<3\pi/2$, $K$ is positive.}.  It is shown at figure (\ref{Figure2}) how  the time evolution of the mean-square displacement depends on the initial position of the particle on the Torus. Clearly, the particle`s diffusion decelerates faster at the region where $K<0$  than the way how the particle`s diffusion proceeds at the region where $K>0$. On the contrary, at the parallels ($K=0$) the particle`s diffusion accelerates;  in fact, the MSD in the parallels is given by
\begin{eqnarray}
\left<s^{2}\right>_{t}=4Dt+\frac{16}{15 a^2 r^2 }(Dt)^3.
\end{eqnarray} 
In general, the biological membranes have a  wide range of morphologies; from discoidal shapes to catenoidal ones \cite{Lipowsky}.  Also, we can consider wavy surface like membranes with microvilli \cite{Aizenbud}, elliptic paraboloid, hyperbolic paraboloid \cite{Tomoyoshi} and periodic nodal surfaces \cite{Holyst}, where the diffusion is clearly affec\-ted by the geometry. The curvature effects on the diffusion on these surface can be  quantified using equation  (\ref{MSD2}).

\section{Conclusions  and perspectives}

In this paper,  we have studied the brownian motion over a d-dimensional curved manifold. The geodesic distance is used as the displacement of a particle diffusing on the space. Using a Riemann normal frame  we have derived a general formula for the mean-square displacement at the short-time regime. This formula reproduces the standard result for very short-times \cite{Einstein}  and it is given by $O(d)$ isometric covariant terms depending upon the Riemann curvature tensor. In particular, we have explored the diffusion over constant curvature manifolds, where we have  shown that diffusion accelerates for the hyperbolic case, whereas it decelerates for  the spherical one. In addition, we have discussed the two dimensional case, as it is relevant for the diffusion on biological membranes. In particular, the behaviour of the diffusion on a Torus can be classified according to the region (defined by $K<0$, $K>0$ or $K=0$) where the particle starts to move.

As we have mentioned, there are several physical observables that describe the motion of a particle diffusing on a curved space. These observables quantify the curvature effects in different ways.  For instance, as it was discussed at \cite{Faraudo} the mean-square geodesic distance has null curvature effects for the diffusion on developable surfaces, whereas using the parametrization displacement \cite{Holyst} it is clear that there are curvature effects. Indeed,  different physical observables have different manifestations of the  same phenomenon. It is then a rather natural question to ask what is the geometrical and physical content of these  physical observables for the brownian motion on curved manifolds.  

In a different direction, we can explore the large-time regime for the mean-values for  the case of  compact manifolds. In this particular case, the expectation values will have a bounded above limit as a consequence of the  compactness of the manifold \cite{Grygorian}.

\section{Acknowledgments}

The author would like to thank Ramon Casta\~neda Priego and Sendic Estrada Jim\'enez for many  valuable discussions.  The support by PROMEP/103á5/08/3291 grant is acknowledged.

\section{Appendix A. Riemann normal coordinates} 
\subsection{ The gauge condition}
As it is mentioned at \cite{Muller}  the second condition of (\ref{defRNC}) is equivalent to the gauge condition (\ref{normality}). In order to prove this we use the Christoffel symbols, therefore we have
\begin{eqnarray}
y^{a}y^{b}\Gamma^{c}_{~ab}(y)=\frac{1}{2}g^{ce}\left(2y^{a}y^{a}\partial_{a}g_{eb}-y^{a}y^{b}\partial_{e}g_{ab}\right).
\label{proof}
\end{eqnarray}
Now, the first derivative on the second condition of (\ref{defRNC}) is $g_{ab}+y^{e}\partial_{a}g_{eb}=\delta_{ab}$, then  $y^{a}y^{b}\partial_{a}g_{eb}=0$. Similarly, the second term of (\ref{proof}) vanishes.

\subsection{An affine connection expansion}
In order to derive (\ref{MetricRNC}) we follow the same procedure presented at \cite{Denjoe}.  Let $\omega$ be a matrix  one-form connection given by $\omega=\omega_{a}dx^{a}$,  where the matrix one-form component $(\omega_{a})^{b}_{c}=\Gamma^{b}_{~ ac}$ is given by the Christoffel symbols. Let us consider the Lie derivative $\mathcal{L}_{X}$ along the radial vector $X=y\cdot\partial$ acting on $\omega$, 
\begin{eqnarray}
\mathcal{L}_{X}\omega=\left[(1+y\cdot\partial)\omega_{a}\right]dy^{a},
\label{eq1}
\end{eqnarray}
where $y^{a}$ are the Riemann normal coordinates. The Lie derivative can be written as
\begin{eqnarray}
\mathcal{L}_{X}=di_{X}+i_{X}d,
\label{identity}
\end{eqnarray}
where $d$ and $i_{X}$ are the exterior and interior derivative, respectively \cite{Nakahara}. Observe,  that the gauge condition (\ref{normality}) can be written as $i_{X}\omega=0$. Therefore using the Cartan structure equation for the curvature, $R=d\omega+\omega\wedge\omega$, is easy to get
\begin{eqnarray}
\mathcal{L}_{X}\omega&=&i_{X}R-i_{X}\left(\omega\wedge\omega\right)\nonumber\\
&=&i_{X}R\nonumber\\
&=&y^{a}{\mathcal R}_{ab}dy^{b}.
\label{eq2}
\end{eqnarray}
Equating (\ref{eq1}) and (\ref{eq2}), we get the condition  $(1+y\cdot\partial)\omega_{a}=y^{a}{\mathcal{R}}_{ab}$, where $({\mathcal R}_{ab})^{c}_{~d}=R^{c}_{~dab}$ is the Riemann curvature tensor . A Taylor expansion centered at  $0$ in both sides of this condition gives
\begin{eqnarray}
\Gamma^{a}_{~bc}=\sum^{\infty}_{k=0}\frac{\left(y\cdot\partial\right)^{k}}{k!\left(k+2\right)}y^{d}R^{a}_{~cdb},
\end{eqnarray}
where Riemann curvature tensor is evaluated at $0$. 

\subsection{A vielbein expansion}
\noindent As it is point out in \cite{Denjoe}, it is sufficient to compute the Taylor expansion of the vielbein $\theta=\theta_{a}dx^{a}$ in order to find (\ref{MetricRNC}). $\theta$ satisfy the free torsion condition
$d\theta+\omega\wedge\theta=0$ and the metric can be re-written in terms of the vielbein as  $g_{ab}=\theta^{i}_{~a}\theta^{j}_{~b}\delta_{ij}$.  Using (\ref{identity}) and the gauge condition, the Lie derivative on $\theta$  is given by
\begin{eqnarray}
\mathcal{L}_{X}\theta=(i_{X}\theta)\omega+di_{X}\theta.
\label{1Lie}
\end{eqnarray}
As $\theta$ is one-form, $i_{X}\theta$ is scalar.  Therefore the second Lie derivative of last expression is
\begin{eqnarray}
\mathcal{L}_{X}\mathcal{L}_{X}\theta=X\left[i_{X}\theta\right]\omega+\left(i_{X}\theta\right)\mathcal{L}_{X}\omega+\mathcal{L}_{X}di_{X}\theta .
\label{2Lie} 
\end{eqnarray}
 The additional gauge condition  $y^{a}g_{ab}\left(y\right)=y^{a}\delta_{ab}$  is equivalent to choose $y^{a}\theta^{i}_{a}\left(y\right)=y^{a}\delta^{i}_{a}$, then  $X\left[i_{X}\theta\right]=i_{X}\theta$ and $\mathcal{L}_{X}di_{X}\theta=di_{X}\theta$. Combining equation (\ref{1Lie}) with (\ref{2Lie}), we get $\mathcal{L}_{X}\left(\mathcal{L}_{X}-1\right)\theta=\left(i_{X}\theta\right)\mathcal{L}_{X}\omega$. Now using equation (\ref{eq2}), we find
 \begin{eqnarray}
\mathcal{L}_{X}\left(\mathcal{L}_{X}-1\right)\theta=\left(i_{X}\theta\right)i_{X}R.
\end{eqnarray}
The condition for the vielbein is then given by $ \left(y\cdot\partial+1\right)(y\cdot\partial)\theta^{i}_{a}=y^{j}y^{b}R^{i}_{~jbc}\theta^{c}_{a}$ . By Taylor expansion  centered at $0$ in both sides of this condition we get
\begin{eqnarray}
\theta^{i}_{a}&=&\delta^{i}_{a}+\sum^{\infty}_{k=2}\frac{\left(y\cdot\partial\right)^{k-2}\left[R^{i}_{~\cdot\cdot c}\theta^{c}_{a}\right]}{k\left(k+1\right)\left(k-2\right)!},\nonumber\\
y\cdot\partial \theta^{i}_{a}&=&0
\end{eqnarray}
where $R^{i}_{~\cdot\cdot c}\equiv y^{a}y^{b}R^{i}_{abc}$. This system of equations can be solved iteratively and they give the following expression for the vielbein until fourth order in $y^{a}$
\begin{eqnarray}
\theta^{i}_{~a}=\delta^{i}_{a}+\frac{1}{3}R^{i}_{~\cdot\cdot a}+\frac{1}{12}y\cdot \partial R^{i}_{~\cdot\cdot a}+\frac{1}{40}\left(y\cdot \partial\right)^{2}R^{i}_{~\cdot\cdot a}+\frac{1}{120}R^{i}_{~\cdot\cdot c}R^{c}_{~\cdot\cdot a}+\cdots
\end{eqnarray}
The metric (\ref{MetricRNC}) is then obtained using $g_{ab}=\theta^{i}_{~a}\theta^{j}_{~b}\delta_{ij}$. 

\subsection{Determinant of the metric}

A Taylor expansion of the determinant of the metric $g\equiv \det g_{ab}$ can be obtained using the identity $\log \det g_{ab}=tr\log g_{ab}$. Since the metric (\ref{MetricRNC}) can be written as $g_{ab}=\delta_{ab}+\Lambda_{ab}\left(y\right)$, its logarithm is given by $\log(1+\Lambda)\approx \Lambda-\frac{1}{2}\Lambda^{2}$.  The trace of this term is then given by
\begin{eqnarray}
-\log g&=&\frac{1}{3}R_{ab}\left(0\right)y^{a}y^{b}+\frac{1}{6}\nabla_{e}R_{ab}\left(0\right)y^{a}y^{b}y^{e}+\frac{1}{90}R^{a}_{cdf}\left(0\right)R^{f}_{gha}\left(0\right)y^{c}y^{d}y^{g}y^{h}\nonumber\\&+&\frac{1}{20}\nabla_{e}\nabla_{f}R_{ab}\left(0\right)y^{a}y^{b}y^{e}y^{f}+\cdot\cdot\cdot .\nonumber\\
\end{eqnarray}

\subsection{Geometric  $G_{3}$ factor}

In order to get the geometric factor $G_{3}\equiv \Delta_{g}^{3}s^{2}$, we need at least the Taylor expansion of $\Delta^{2}_{g}s^{2}$ until second order $O(y^{2})$. By straighforward calculation this expansion is given by

\begin{eqnarray}
\Delta^{2}_{g}s^{2}&=&-\frac{4}{3}R_{g}-y^{a}\partial_{a}R_{g}-2y^{a}\nabla^{b}R_{ab}+
\frac{4}{3}R^{a}_{c}R_{ad}y^{c}y^{d}+\frac{4}{9}\left.R^{a}_{~cd}\right.^{b}R_{ab}y^{c}y^{d}-\frac{4}{45}\left\{-2R^{a}_{f}R^{f}_{~gha}y^{g}y^{h}\right.\nonumber\\&+&\left. R^{ag}_{~~df}R^{f}_{~gha}y^{d}y^{h}+R^{ah}_{~~df}R^{f}_{~gha}y^{d}y^{g}+R^{a}_{~cgf}R^{fg}_{~~ha}y^{c}y^{h}+R_{ac}~^{hf}R_{fgh}~^{a}y^{c}y^{g}\right\}
-\frac{2}{5}\left(\nabla^{2}R_{cd}\right)y^{c}y^{d}\nonumber\\&-&\frac{2}{5}\left(\nabla_{e}\nabla_{f}R_{g}\right)y^{e}y^{f}-\frac{8}{5}\left(\nabla^{e}\nabla_{f}R_{ed}\right)y^{f}y^{d}+O(y^{3}).\end{eqnarray}

\section{Appendix B. $\Delta_{g}$ on  a scalar function} Let $\Omega:\mathbb{M}\to \mathbb{R}$ be a scalar differentiable function on the manifold. Let us take a Riemann normal system of coordinates centered at $y^{a}=0$, then the Laplace-Beltrami operator  acting on $\Omega$ is given by the second derivative of $\Omega$, i.e., 
\begin{eqnarray}
\left.\Delta_{g}\Omega\right|_{y=0}=\left.\partial^{2}\Omega\right|_{y=0}.
\end{eqnarray}
In order to show this result we split the inverse metric and the square root of the metric determinant as
\begin{eqnarray}
g^{ab}&=&\delta^{ab}+\Lambda^{ab}\left(y\right)\nonumber\\
\sqrt{g}&=&1+\lambda\left(y\right),
\end{eqnarray}
where $\Lambda^{ab}\left(y\right)$ and $\lambda\left(y\right)$ can be found in terms of covariant derivative of the Riemann curvature as we have seen above.  This functions satisfy the following properties: $\Lambda^{ab}\left(0\right)=0$, $\partial_{c}\Lambda^{ab}\left(0\right)=0$, $\lambda\left(0\right)=0$, and  $\partial_{c}\lambda\left(0\right)=0$. Thus the action of $\Delta_{g}$ on $\Omega$ can be written as
\begin{eqnarray}
\Delta_{g}\Omega&=&\left\{\partial^{2}\Omega+\partial_{a}\lambda\left(y\right)\partial^{a}\Omega+\lambda\left(y\right)\partial^{2}\Omega+\partial_{a}\Lambda^{ab}\left(y\right)\partial_{b}\Omega+\Lambda^{ab}\left(y\right)\partial_{a}\partial_{b}\Omega\right.	\nonumber\\
&+&\left.\left.\partial_{a}\lambda\left(y\right)\Lambda^{ab}\left(y\right)\partial_{b}\Omega+\lambda\left(y\right)\partial_{a}\Lambda^{ab}\left(y\right)\partial_{b}\Omega+\lambda\left(y\right)\Lambda^{ab}\left(y\right)\partial_{a}\partial_{b}\Omega\right\}\right|_{y=0}\nonumber\\
\end{eqnarray}
Then after the substitution $y=0$, it is obtained the desired result.


\begin{thebibliography}{99}

\bibitem{Colloid} Jan K.G. Dhont, An Introduction to Dynamics of Colloids, 1996 Elsevier Science.
\bibitem{Duplantier1} Bertrand Duplantier,{\it Phys. Rev. Lett.} {\bf 81} (1998). 
\bibitem{Almeida} P.F.F. Almeida and W.L.C. Vaz. Handbook of Biological Physics (Elsevier Science, Amsterdam, 1995), Vol. 1, 
Chap. 6.
\bibitem{Weiss} Matthias Weiss, Hitoshi Hashimoto, and Tommy Nilsson, {\it    Biophys. J.} {\bf 84} pp.  4043Ð4052 ( 2003)
\bibitem{Ivo} Ivo F. Sbalzarini, Arnold Hayer, Ari Helenius, and Petros Koumoutsakos, {\it Biophys. J.} {\bf 90} (2006).
\bibitem{Seifert}  E. Reister and U. Seifert, {\it Europhys. Lett.} {\bf 71} pp. 859-865, 
(2005).  cond-mat/0503568. 
\bibitem{Naji} Ali Naji and Frank L. H. Brown, {\it J. Chem. Phys.} {\bf 126} p. 235103 (2007). 
\bibitem{Seifert2} Ellen Reister-Gottfried, Stefan M. Leitenberger, and Udo Seifert, {\it Phys. Rev. E} {\bf 81} p. 031903 (2010). 
\bibitem{Aizenbud} Boris M. Aizenbud and Nahum D. Gershon, {\it Biophys. J.} {\bf 38}, 287 (1982).
\bibitem{Gustaffson}  S. Gustafsson and B. Halle, {\it J. Chem. Phys.}  {\bf 106}, 1880 (1997).
\bibitem{H}  D. Anderson and H. Wennerstr$\ddot{o}$m, {\it J. Phys. Chem.} {\bf 94}, p. 8683 (1990).
\bibitem{K} J. Balakishnan, {\it Phys. Rev. E} {\bf 61} 4648 (2000)
\bibitem{Holyst} R. Holyst, D. Plewczynski, and A. Aksimentiev, {\it Phys. Rev. E} {\bf 60} 302 (1999).
\bibitem{Christensen} Micheal Christensen, {\it Journal of Computational Physics} {\bf 201} pp. 421-438 (2004).
\bibitem{Kac} Mark Kac {\it The American Mathematical Monthly}, {\bf  73}, pp. 1-23 (1966)
\bibitem{Einstein} A. Einstein, {\it Ann. Phys.} (Leipzig) {\bf 17}, 549 (1905).
\bibitem{Faraudo} Jordi Faraudo, {\it J. Chem. Phys.} {\bf 116} 5831 (2002).
\bibitem{Tomoyoshi} Tomoyoshi Yoshigaki, {\it Phys. Rev. E} {\bf 75} 041901 (2007).
\bibitem{Fabrice} F. Debbasch and M. Moreau.  {\it Phys. A}, 343:81, 2004. 
\bibitem{Castro} P. Castro-Villarreal, on preparation preprint.
\bibitem{Eisenhart} Luther Pfahler Eisenhart, Riemannian Geometry. Princeton University Press (1997).
\bibitem{Denjoe} Denjoe O«Connor, Quantum field theory in curved spacetime: the effective action and finite size effects. {\it PhD thesis} University of Maryland (1985).
\bibitem{Denjoe1} Piotr Amsterdamski and Denjoe O' Connor, {\it Nuclear Physics B} {\bf 298} pp. 429-444 (1988).
\bibitem{Denjoe2} Piotr Amsterdamski, Andrew L. Berkin and Denjoe O' Connor, {\it Class. Quantum Grav.} {\bf 6} pp. 1981-1991 (1989).
\bibitem{Muller} Uwe M$\ddot{u}$ller, Christian Schubert and Anton E. M. van de Ven {\it Gen. Rel. and Grav.} {\bf 31} p. 1759 (1999)
\bibitem{Hatzinikitas} A Hatzinikitas - Arxiv preprint hep-th/0001078
\bibitem{Nakahara} Mikio Nakahara, Geometry, Topology and Physics. IoP Publishing (2003).
\bibitem{DeWitt} Bryce S. DeWitt, Dynamical Theory of Groups and Fields, Published by Gordon Breach (1965) (New York).
\bibitem{Grygorian} Alexander GrigorÕyan,  Estimates of heat kernels on Riemannian manifolds,  Notes (website).
\bibitem{spivak} Spivak M. A., A Comprehensive Introduction to Differential Geometry. Vol. Four, Third  Edition, Publish or Perish, Inc. (1999). 
\bibitem{Lipowsky} Reinhard Lipowsky, Nature, Vol. 349, p. 475 (1991). 
\end{thebibliography}
\end{document}